\documentclass[prl,aps,twocolumn,showpacs,superscriptaddress]{revtex4-1}

\usepackage{graphicx}
\usepackage{dcolumn}
\newcolumntype{d}{D{.}{.}{2}}
\usepackage{bm}
\usepackage{float}
\usepackage{amsmath}
\usepackage{amssymb}

\begin{document}


\title{Stable all-nitrogen metallic salt at terapascal pressures}

\author{Jian Sun}
\email[]{To whom correspondence should be addressed. E-mail: jiansun@nju.edu.cn}
\affiliation{Department of Physics and National Laboratory of 
Solid State Microstructures, Nanjing University, Nanjing 210093, China}
\affiliation{Lehrstuhl f\"ur Theoretische Chemie, Ruhr-Universit\"at
  Bochum, 44780 Bochum, Germany}
\affiliation{Theory of Condensed Matter Group,
Cavendish Laboratory, J J Thomson Avenue, Cambridge CB3 0HE,
  United Kingdom}

\author{Miguel Martinez-Canales}
\email[]{To whom correspondence should be addressed. E-mail: miguel.c.martinez@ucl.ac.uk}
\affiliation{Department of Physics and Astronomy,
University College London, Gower Street, London WC1E 6BT, United Kingdom}

\author{Dennis D.\ Klug}
\affiliation{Steacie Institute for Molecular
  Sciences, National Research Council of Canada, Ottawa, K1A 0R6,
  Canada}

\author{Chris J.\ Pickard}
\affiliation{Department of Physics and Astronomy,
University College London, Gower Street, London WC1E 6BT, United Kingdom}

\author{Richard J.\ Needs} 
\affiliation{Theory of Condensed Matter Group, 
Cavendish Laboratory, J J Thomson Avenue, Cambridge CB3 0HE,
  United Kingdom}


\date{\today}

\begin{abstract}
  The phase diagram and equation of state of dense nitrogen are of
  interest in understanding the fundamental physics and chemistry
  under extreme conditions, including planetary processes, and in
  discovering new materials.
  We predict several stable phases of nitrogen at multi-TPa pressures,
  including a $P4/nbm$ structure consisting of partially charged
  N$_{2}^{\delta+}$ pairs and N$_{5}^{\delta-}$ tetrahedra, which is
  stable in the range 2.5--6.8 TPa.
  This is followed by a modulated layered structure between 6.8 and
  12.6 TPa, which also exhibits significant charge transfer.
  The $P4/nbm$ metallic nitrogen salt and the modulated structure
  are stable at high pressures and temperatures, and
  they exhibit strongly ionic features and charge density distortions,
  which is unexpected in an element under such extreme conditions and
  could represent a new class of nitrogen materials.
  The P-T phase diagram of nitrogen at TPa pressures is investigated
  using quasi-harmonic phonon calculations and \textit{ab initio} molecular
  dynamics simulations.

\end{abstract}




\maketitle




%

%
%

%

The polymerization of molecular nitrogen under pressure has been very
actively researched over the past two decades and has stimulated many
experimental \cite{Goncharov2000,Eremets2001,
  Gregoryanz2001,Gregoryanz2002,Eremets2004,Eremets2004a,Eremets2007,
  Gregoryanz2007,Lipp2007,Mukherjee2007,Goncharov2008,Mattson2011} and theoretical
\cite{McMahan1985,Martin1986,Lewis1992,Mailhiot1992,
  Mitas1994,Alemany2003,Mattson2004,Zahariev2005,Oganov2006a,Uddin2006,Zahariev2006,
  Caracas2007,Wang2007a,Zahariev2007,Yao2008,Pickard2009-nitrogen,Ma2009-nitrogen,Boates2011,Wang2012-diamondoid-N}
investigations.  Density functional theory (DFT) studies suggested
that dissociation of nitrogen could occur at high pressures which
are, nevertheless, attainable in diamond anvil cell (DAC)
experiments.  In a ground-breaking paper, Mailhiot \textit{et al.}\
\cite{Mailhiot1992} predicted polymerization of nitrogen molecules
under pressure, leading to the formation of the ``cubic gauche'' 
(\emph{cg}) framework structure (space group $I2_{1}3$).  
After considerable efforts
\cite{Goncharov2000,Eremets2001}, \emph{cg} nitrogen was finally
synthesized by Eremets \textit{et al.}\ \cite{Eremets2004}, a decade
after its prediction.  
Novel techniques must be developed if \emph{cg}-N is to be recovered to
ambient pressure \cite{Eremets2004}, which could lead to the synthesis
of polymeric nitrogen \cite{Eremets2004} or high--N content salts
\cite{Haiges2004}.  Such materials are expected to be excellent
candidates as high-energy-density materials.

The N$\equiv$N triple bond is one of the strongest known, and breaking
it requires surmounting a substantial energetic barrier.  However, at
high temperatures the triple bond breaks at pressures above about 110
GPa \cite{Eremets2004}, a much lower pressure than predicted for the
single bond in H$_2$ \cite{Pickard2007-hydrogen} and the double bond
in O$_2$, which is predicted to survive up to 2 TPa
\cite{Sun2012-O2,Zhu2012}. Once the N$\equiv$N triple bond is broken,
a wide variety of structures can be adopted, similar to phosphorus and
arsenic. 

After extensive searches for high-pressure nitrogen structures, the
sequence of low-temperature phase transitions \emph{cg} $\rightarrow
Pba2 \rightarrow P2_12_12_1$ appeared to be accepted
\cite{Ma2009-nitrogen,Pickard2009-nitrogen}. That view was challenged
recently, when a cagelike diamondoid ``N10'' structure 
(space group $I\bar{4}3m$) was found to be
more stable than any other candidate above 263 GPa
\cite{Wang2012-diamondoid-N}.
Recently, the melting behavior of nitrogen and phases beyond
\emph{cg}-N have also been
investigated \cite{Mukherjee2007,Goncharov2008,Mattson2011,Boates2011}.

Recently, dynamical shock wave \cite{Jeanloz2007,Knudson2008,Eggert2010} and
ramped compression experiments \cite{Hawreliak2007,Bradley2009,rygg2012} have
increasingly been used to investigate materials at TPa pressures.
Even more extreme conditions are attainable today in laser ignition
experiments \cite{Kritcher2011} and laser-induced microexplosions
\cite{Mizeikis2012}.
Experimental determinations of structures at TPa pressures will soon
be possible \cite{rygg2012}, and experiments aimed at compressing
nitrogen to TPa pressures are ongoing \cite{Millot-pc}.  However,
there is very little knowledge of the structures and properties of
nitrogen under these conditions, and therefore theoretical predictions
are particularly important
\cite{Sun2009-carbon,McMahon2011,Pickard2010-Al,Martinez2012-carbon,Sun2012-O2}.
Understanding the energetics of elements is crucial before one can
understand the energetics of the compounds they might form.

Although an increase in coordination number with pressure seems
physically reasonable, it is by no means universal.
Aluminum, for example, is expected to adopt more open structures at TPa
pressures as the valence electrons move away from the ions in the
formation of ``electride structures'' \cite{Pickard2010-Al}.
Well-packed structures are strongly disfavored in oxygen up to at
least 25 TPa \cite{Sun2012-O2}, which is well beyond dissociation.
The coordination number of nitrogen has not been determined beyond 800
GPa, at which pressure the insulating N10 ($I\bar{4}3m$) phase is stable.



\begin{figure}[tbh]
\begin{center}
\includegraphics[width=0.45\textwidth]{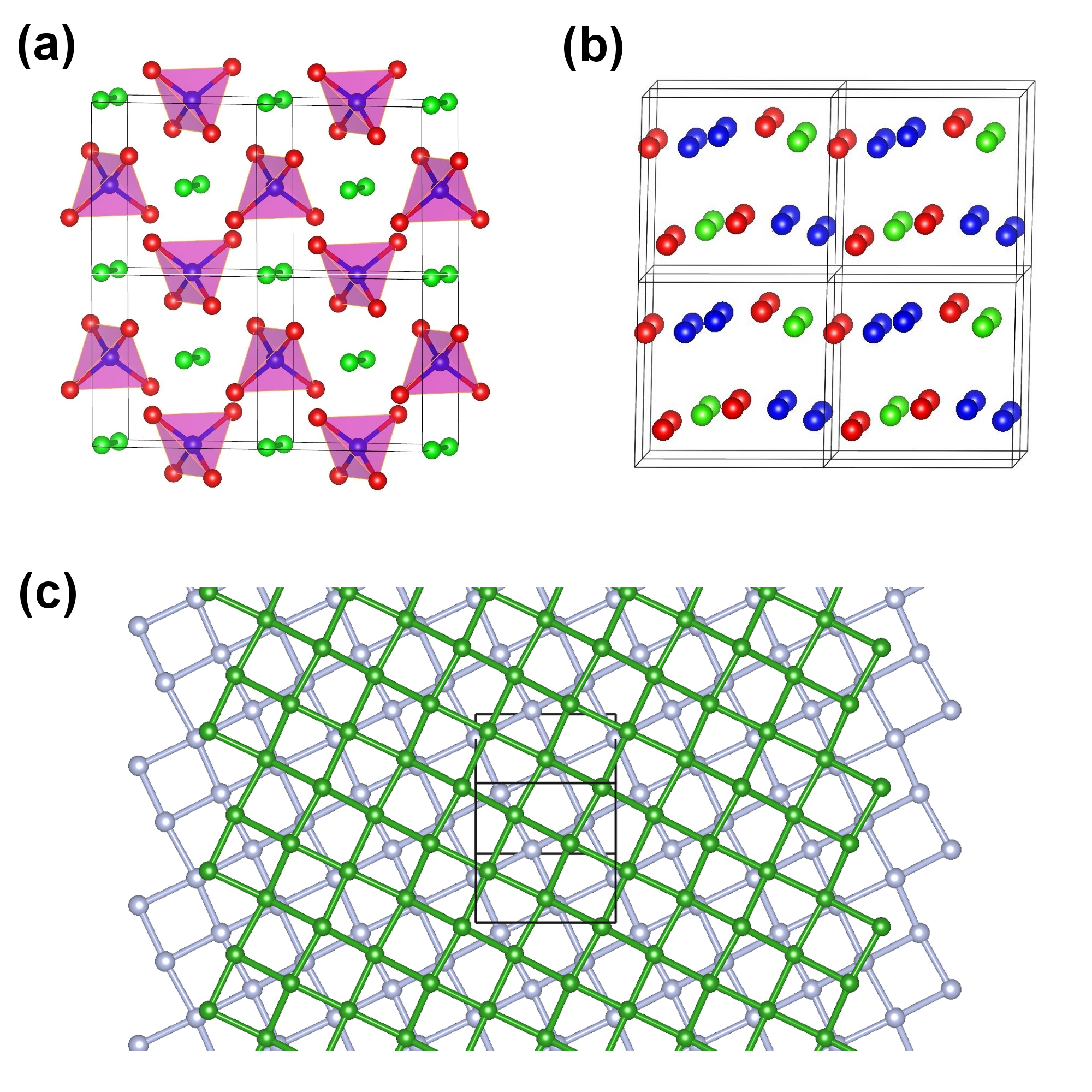}
\caption{%
  (color online). Crystal structures of the newly predicted nitrogen
    phases.  
  (a) $P4/nbm$, (b) $P2_1$, (c) a supercell of $P2_1$ viewed along the
  $a$-axis.  
  The colored spheres in (a) and (b) represent atoms with
  different charge environments, as described in the text. 
  Atoms in different layers in (c) have been shown with different colors.
}
\label{fig:lattice}
\end{center}
\end{figure}

We have used \textit{ab initio} random structure searching (AIRSS)
\cite{Pickard2006-silane,Pickard2011-review} and DFT methods to find
candidate structures of nitrogen up to multi-TPa pressures.
Details of the searches are provided in the Supplemental Material \cite{epaps}.
These searches have enabled us to identify a number of novel candidate
structures.
Four of them are thermodynamically stable within certain pressure
ranges, namely, $P4/nbm$ ($Z$=14), where $Z$ is the number of atoms in
the unit cell, $P2_1$ ($Z$=10), $R\bar{3}m$
($Z$=3) and $I4_1/amd$ ($Z$=4).
The most important structures are shown in Fig.\ \ref{fig:lattice} 
and defined in Table \ref{table:charge}.
Most interestingly, the tetragonal $P4/nbm$ structure is characterized
by the presence of N$_2$ pairs and N$_5$ tetrahedra.
At 2.5 TPa, the N-N bond length within the N$_5$ tetrahedra is 1.13
\AA, and the shortest distance between the corners of the tetrahedra
and N$_2$ dimers is 1.26 \AA.  The N-N bond length within the dimer of
about 1.17 \AA, which is significantly shorter than the N-N separation
between adjacent dimers (about 1.29 \AA), implying that the dimers are
separated.

\begin{table*}[tp]
\setlength{\tabcolsep}{5pt}
\caption{Structures and Bader charges ($Q_B$) of the newly predicted stable nitrogen phases.
}                   
\centering
\begin{tabular}{l c l r l d}
\hline\hline                                        
Structures    &$P$ (TPa) &Lattice parameters (\AA)   & \multicolumn{2}{l}{Atomic positions}  &\multicolumn{1}{c}{$Q_B$} ($|e|$)   \\ [0.5ex] 
\hline
$Cmca$        & 2.5    &$a$=2.090, $b$=3.041, $c$=2.617      & 8f & (0.0,  0.3966,  0.8157)     &0   \\[+0.1cm]        
$P4/nbm$      & 2.5    &$a$=3.424, $c$=2.466      & 2d & (0.0, 0.5, 0.5)             &-0.07   \\    
&        &    & 4g & (0.0, 0.0, 0.7376)          &0.18      \\
&        &                                     & 8m & (0.1808, 0.6808, 0.7879)    &-0.07    \\[+0.1cm]
$P2_1$ & 8.0 & $a=2.579$, $b=2.396$, $c=2.390$          & 2a & (0.7889, 0.9079, 0.0325) & 0.03 \\
	 & 	& $\beta = 62.903^{\circ}$ 		& 2a & (0.2912, 0.2331, 0.5345) & 0.01 \\
	 & 	& 					& 2a & (0.1858, 0.6314, 0.3824) & -0.20 \\
	 & 	&  					& 2a & (0.2487, 0.8206, 0.7525) & 0.35 \\
 	 & 	& 					& 2a & (0.6866, 0.5078, 0.8789) & -0.19 \\[+0.1cm]
$R\bar{3}m$   & 12     &$a$=1.613, $c$=1.468      & 3b & (0.0, 0.0, 0.5)  &0   \\[+0.1cm]
$I4_1/amd$    & 30     &$a=0.890$, $c=3.452$       &4a & (0.0, 0.25, 0.875)  &0   \\
\hline\hline                                        
\end{tabular}
\label{table:charge}                                
\end{table*}

The unique structure of the $P4/nbm$ phase led us to perform a Bader
charge transfer analysis \cite{Bader}.
As summarized in Table\ \ref{table:charge}, we found that the N$_5$
tetrahedra are negatively charged ($-0.37 |e|$) and serve as anions,
while the positively charged N$_2$ dimers ($+0.37 |e|$) act as
cations.
The charge transfers depends on the precise definition of the atomic
charges, but the computed charges are substantial, and the overall
picture is independent of their definition.
We therefore suggest that the $P4/nbm$ structure resembles an
all-nitrogen salt.

Our best candidate structure beyond 6.8 TPa, of symmetry $P2_1$ with
10 atoms in the unit cell, also shows very strong charge transfer. The
computed Bader charges range between -0.20 and +0.35 $|e|$.
This structure can be viewed as two distorted and buckled square
layers, rotated with respect to one another by about $\arctan
\frac{3}{4}$.  Similar energetically competitive structures,
such as the $Fdd2$ structure, can be formed with different rotation
angles.
It is likely that the most stable structure
at these pressures is an incommensurate charge density wave (CDW)
phase.  Nevertheless, even after additional directed searches as
described in the Supplemental Material \cite{epaps}, the 10-atom
$P2_1$ phase remains the best candidate.  The appearance of strong
charge transfer effects in an element at such high pressures is most
unexpected, as the resulting Coulomb energy is substantial.

At lower pressures, we found a $Cmca$ structure similar to ``black
phosphorus'' to be stable within a narrow pressure range.  
As depicted in the Supplemental Material \cite{epaps}, the $Cmca$
structure consists of three-fold-coordinated nitrogen atoms and has
zig-zag layers, with a shortest N-N distance of about 1.15 \AA\ at 2.5
TPa, which is much shorter than the N-N separation between the layers
of about 1.56 \AA.
Layered structures also appear in the high-pressure phases of other
small molecules of first row atoms, e.g., CO \cite{Sun2011-CO} and
CO$_2$ \cite{Sun2009-CO2}.
%

\begin{figure}[th]
\begin{center}
\includegraphics[width=0.45\textwidth]{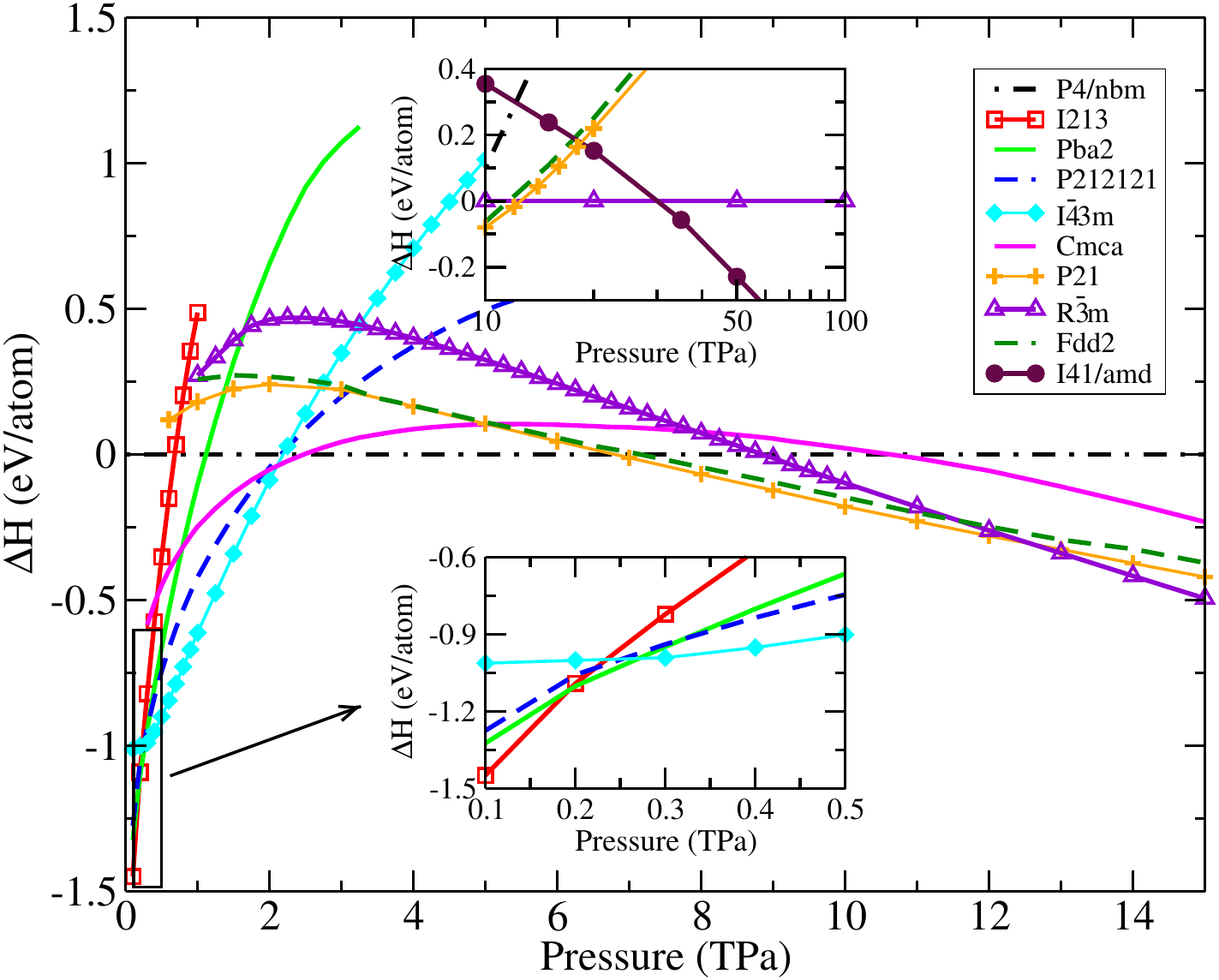}
\caption{%
  (color online). Enthalpy-pressure relation for N. The values in the main figure are relative
  to the enthalpy of $P4/nbm$. The upper inset shows the enthalpies 
  relative to $R\bar{3}m$ at very high pressures. The lower inset shows
  the enthalpies at low pressures relative to $P4/nbm$.
  }
\label{fig:eos}
\end{center}
\end{figure}

As can be seen in the enthalpy-pressure relations of Fig.\
\ref{fig:eos}, solid nitrogen undergoes a series of structural phase
transitions beyond the previously-known polymeric phases:
$I\bar{4}3m \xrightarrow{\rm 2.1\ TPa} Cmca \xrightarrow{\rm 2.5\ TPa}
P4/nbm \xrightarrow{\rm 6.8\ TPa} P2_1 \xrightarrow{\rm 12.6\ TPa}
R\bar{3}m \xrightarrow{\rm 30\ TPa} I4_1/amd$.
The partially ionic $P4/nbm$ phase is stable over a wide pressure
range, but compression to about 6.8 TPa leads to the layered $P2_1$
structure.
The $P2_1$ structure is the most favorable phase in the range 6.8--12.6
TPa, whereupon it transforms into a six-fold-coordinated hexagonal
$R\bar{3}m$ phase.  As shown within the inset to Fig.\ \ref{fig:eos},
nitrogen forms a $I4_1/amd$ structure similar to Cs-IV
\cite{Takemura1982-Cs-IV} at about 30 TPa.

We have established the dynamical stability of the proposed structures
by computing their phonon spectra over a wide range of pressures.  A
selection of phonon spectra is shown in the Supplemental Material
\cite{epaps}.  The phonon dispersion relation of $P4/nbm$ shows steep
acoustic branches together with fairly flat optical modes.  This
supports the view that $P4/nbm$ is formed of ``units''; the N--N
vibrons as well as the breathing mode of the tetrahedra are
essentially dispersionless, and are well screened by the electronic cloud.
One might expect $P2_1$ to show two relatively soft modes
corresponding to shearing of layers but, in fact, while one of them
drops to a very low frequency of 187 cm$^{-1}$ at $\Gamma$, the other
has a much higher frequency of about 1210 cm$^{-1}$.
This difference in frequency arises from the alignment of shear directions 
with respect to the ripples in the layers.

\begin{figure}[t]
\begin{center}
\includegraphics[width=0.475\textwidth]{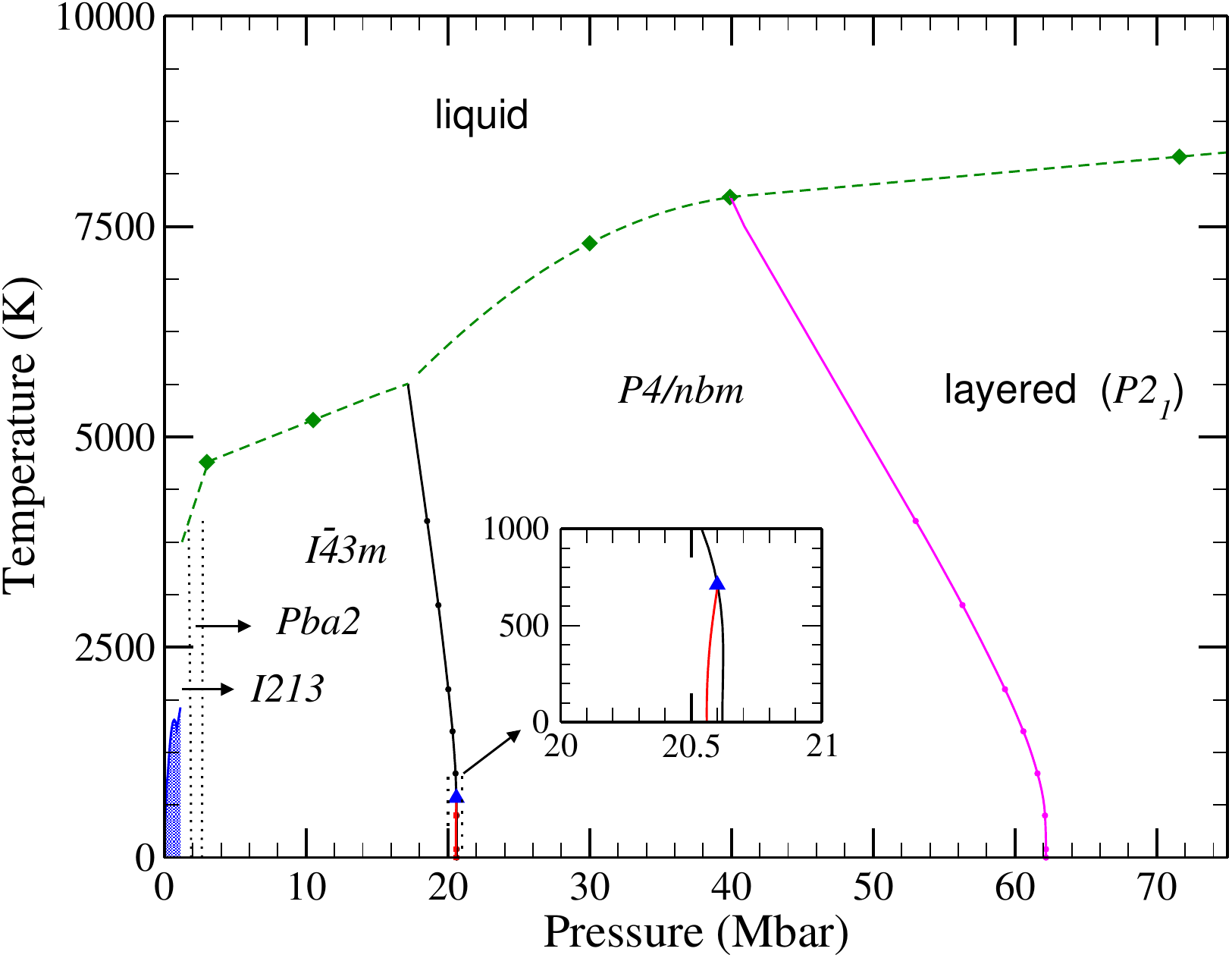}
\caption{%
  (color online). Proposed P-T phase diagram of nitrogen.  The
  solid-solid phase boundaries up to 4000 K were computed using the
  QHA. The melting temperatures were computed using the Z-method and
  \textit{ab initio} molecular dynamics.  The shaded blue area on the
  left corner is the solid molecular N$_2$ region, the inset depicts the
  small stability region of $Cmca$.  }
\label{fig:melting}
\end{center}
\end{figure}

We have extended our calculations to include nuclear motion effects so
that we can investigate their key role in high-temperature dynamical
compression experiments.
At low temperatures and pressures $P < 8$ TPa we have calculated the
P-T phase diagram within the quasiharmonic approximation (QHA).  
We have also investigated higher temperatures using \emph{ab initio}
molecular dynamics, and have calculated a rough melting curve using the
Z--method \cite{Belonoshko2006}
Although the Z-method somewhat overestimates melting temperatures, it
is a straightforward scheme, and provides an estimated upper bound for
the melting temperature, $T_m$.
Our calculated melting temperature could be compared with an
experimental Hugoniot.
Our results suggest a noticeable increase in $T_m$ with pressure,
perhaps monotonic.
The melting temperature is about 4700 K for N10 ($I\bar{4}3m$) 
at 400 GPa, and is as
high as 8700 K at 8 TPa for $P2_1$, which is comparable to carbon
under similar conditions \cite{Correa2008}. 
The phase diagram of Fig.\ \ref{fig:melting} is generated from a 
synthesis of our melting data and QHA vibrational results.

The inclusion of vibrational effects reduces the region of stability
of $Cmca$ to a small wedge in Fig.\ \ref{fig:melting} and, even in
that region, $Cmca$ is never the most stable phase by more than 2
meV/atom.
Because the candidate structures have very different characters, their
respective vibrational energies differ considerably.
As a consequence, vibrational effects make noticeable changes to the
transition pressures even at 0 K where, for example, the pressure for
the $P4/nbm\rightarrow P2_1$ transition is reduced by 0.5 TPa.
The N10 ($I\bar{4}3m$) $\rightarrow P4/nbm$ transition is not affected as much, because 
their enthalpies cross with a much larger gradient.
Finally, our QHA results greatly reduce the possibility of $P2_12_12_1$
being stabilized by temperature.

%


Our calculations show that the insulating behavior of nitrogen
persists to 2.0 TPa, at which the transition to the metallic $Cmca$ and
$P4/nbm$ phases occurs.
This is a considerably higher pressure than in other light elements
such as hydrogen \cite{Pickard2007-hydrogen}, carbon
\cite{Martinez2012-carbon} and oxygen (although oxygen is predicted to
become insulating at 2 TPa \cite{Sun2012-O2}).
The projected density of states (pDOS) shows signs of
$2s\rightarrow 2p$ charge transfer only above 15 TPa.  The electronic band
structures and pDOS of the relevant structures are shown in the
Supplemental Material \cite{epaps}.

\begin{figure}[thp]
\begin{center}
\includegraphics[width=0.50\textwidth]{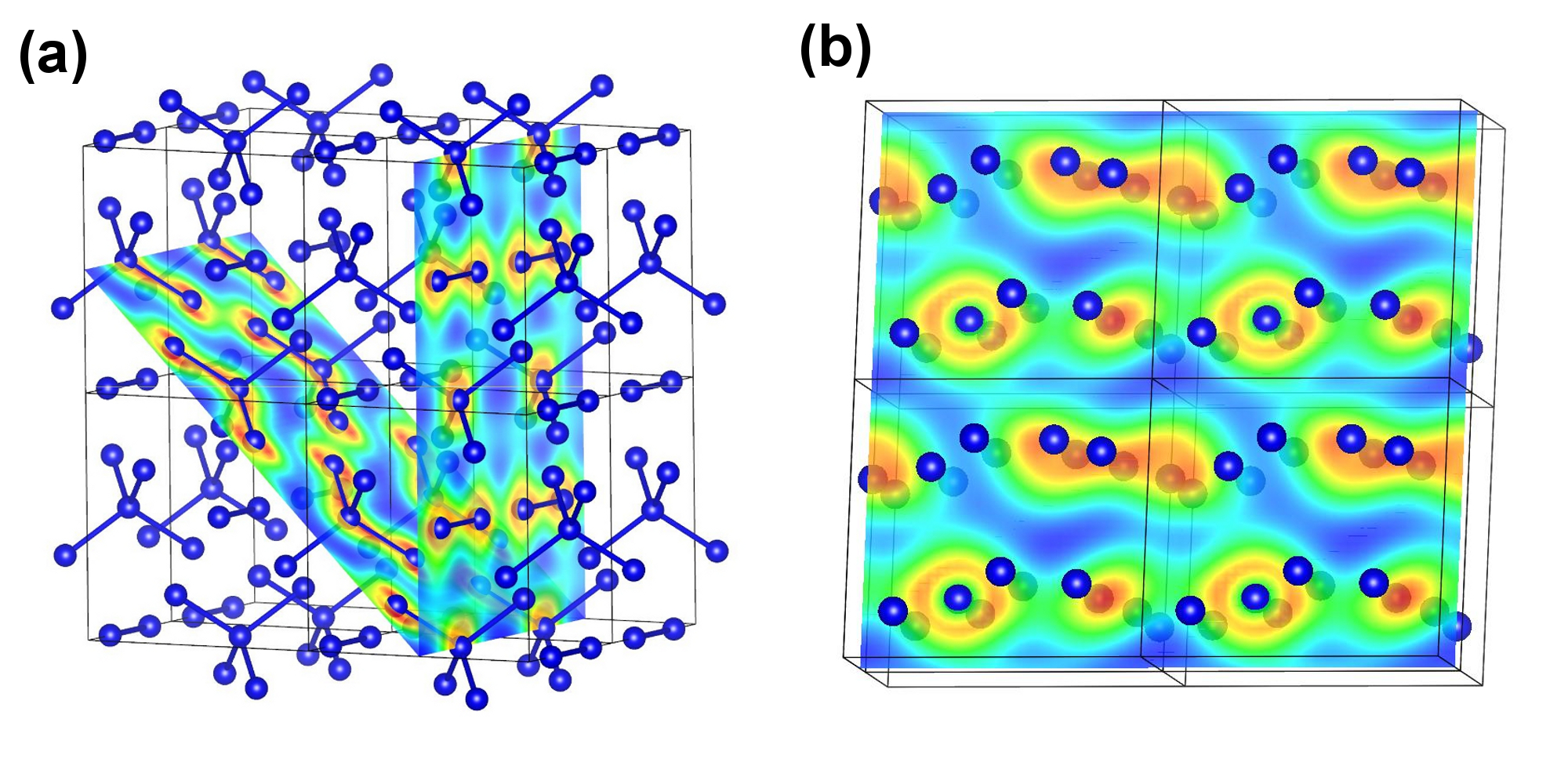}
\caption{%
(color online). Electron densities of polymeric nitrogen.
  (a) $P4/nbm$ at 2.5 TPa along [100] and [110] directions,
  (b) $P2_1$ at 8 TPa along [001] directions.
}
\label{fig:charge_density}
\end{center}
\end{figure}

The formation of N$_2$ pairs and N$_5$ tetrahedra in $P4/nbm$ is
confirmed by the charge density plot in Fig.\
\ref{fig:charge_density}(b), where the atoms between the tetrahedron
center and corners, and between the two atoms within the dimers, form
strong covalent bonds.
The electron localization function (ELF) shown in the Supplemental
Material \cite{epaps} provides additional evidence for the formation
of tetrahedra and dimers in $P4/nbm$.

The emergence of charged units from an elemental compound has also
been observed in $\gamma$--Boron \cite{Oganov2009}.
The pressures of interest in the present study are, however, much
larger than those at which $\gamma$--Boron has been observed and,
besides, $P4/nbm$--nitrogen is a very different system.
The charge transfer in boron satisfies its tendency to form electron
deficient icosahedra.
Based on the structural richness of phosphorus and arsenic one might
assume that nitrogen could form similarly rich bonding patterns once
the triple bond is broken.
At 2 TPa, however, packing efficiency is crucial and one would expect
structures with high coordination numbers to be formed.
In this metallic phase the variation in coordination (or the
inhomogeneity of the density of ions) leads to charge transfer from
the highly-coordinated atoms (``N$_2$ pairs'') to the lower
coordinated ones (``tetrahedra'' corners).
This charge transfer allows the $P4/nbm$ structure to form a unique
metallic all-nitrogen salt.
Each atom in the 1D chains perpendicular to the layers of tetrahedra has
four long bonds to the corners of the tetrahedra, thus only one
electron is left.
The 1D chain with uniform N--N distances is unstable and undergoes a
Peierls distortion, similar to that in lithium \cite{Neaton1999}, and
N$_2$ pairs are formed.

Although $P2_1$ also shows a clear charge transfer, the atomic
arrangement is completely different from that in $P4/nbm$.
This phase is formed by distorted, buckled and rotated square layers, 
resembling a CDW.  
CDWs have been observed previously in chalcogenides under pressure
\cite{Degtyareva2005a}.
We believe that nitrogen may actually adopt an incommensurate structure
between 6 and 12 TPa. 
In light of the charge transfer, the buckling of layers could be
viewed not only as a symmetry breaking, but also as allowing the
positive and negative ions to approach one another and reduce the
energy.
In addition, 
while carbon expels charge from the $2s$ levels above 3 TPa
\cite{Martinez2012-carbon}, eventually forming an electride, the
bottom of the valence band of $P2_1$ still has a large $2s$
population.  Charge depletion from the $2s$ levels only arises with
the transition to the $R\bar{3}m$ phase, at almost 12 TPa.  Electride
structures are found, for example, in carbon
\cite{Martinez2012-carbon} and aluminum \cite{Pickard2010-Al} at very
high pressures, but they do not appear in nitrogen up to at least 100
TPa.  While intuition suggests that close-packed structures should be
favored under extreme pressures, this is not the case for nitrogen
even at 100 TPa where, for example, the face-centered-cubic (fcc)
structure is almost 2 eV per atom higher in enthalpy than $I4_1/amd$.

In conclusion, we have presented a P--T phase diagram for nitrogen
using results from first-principles methods.  We have used the
\emph{ab initio} random structure searching method to investigate 
N at TPa pressures.
We do not find stable close-packed phases below 100 TPa.
We have proposed five additional pressure driven phase transitions and
two novel phases with unusual behavior.  The $P4/nbm$ structure can be
seen as a (N$_2^{\phantom{2}\delta+}$N$_5^{\phantom{2}\delta-}$)$_2$
metallic salt, formed from dimers and tetrahedra.  The layered $P2_1$
structure also shows strong charge transfer and an undulation
resembling a CDW.
Our \emph{ab initio} molecular dynamics and Z--method calculations
provide a first estimate of the melting curve of nitrogen at TPa
pressures.


J.S.\ gratefully acknowledges financial support of the
National Natural Science Foundation of China under
Grants No. 51372112 and No. 11023002, the State Key
Program for Basic Research of China under Grant
No. 2011CB922103, 
the Alexander von Humboldt (AvH) fellowship,
and the Marie Curie fellowship.
M.M.C., C.J.P.\ and R.J.N.\ were supported by the EPSRC.

\end{document}